# Plagiarism deterrence for introductory programming

Simon J. Cohen[+], Michael J. Martin[+], Chance A. Shipley, Abhishek Kumar, Andrew R. Cohen[*]

College of Engineering, Drexel University, Philadelphia PA USA

[+]Authors contributed equally

[*]Correspondence to andrew.r.cohen@drexel.edu

**Abstract**

Plagiarism in introductory programming courses is an enormous challenge for both students and institutions. For students, relying on the work of others too early in their academic development can make it impossible to acquire necessary skills for independent success in the future. For institutions, widespread student cheating can dilute the quality of the educational experience being offered. Currently available solutions consider only pairwise comparisons between student submissions and focus on punitive deterrence. Our approach instead relies on a class-wide statistical characterization that can be clearly and securely shared with students via an intuitive new p-value representing independence of student effort. A pairwise, compression-based similarity detection algorithm captures relationships between assignments more accurately. An automated deterrence system is used to warn students that their behavior is being closely monitored. High-confidence instances are made directly available for instructor review using our open-source toolkit. An unbiased scoring system aids students and the instructor in understanding true independence of effort. Preliminary results indicate that the system can provide meaningful measurements of independence from week one, improving the efficacy of technical education.

**Introduction**

Plagiarism is endemic in early computer science education [1]. For students new to programming, independent development of skills is crucial for long-term success. Existing tools are generally intended to detect plagiarism rather than deter it. Detection typically requires conclusive evidence of plagiarism and results in stringent academic outcomes. Existing detection tools for early programming are often underutilized due to lack of features and usability issues [2]. The deterrence approach presented here is built instead by establishing, for each student submission, the probability of independent effort as measured across the entire class. This system is designed to proactively encourage students to develop skills individually from the very first assignment, helping to build independent coding practices early on needed for long-term success.

Traditionally, plagiarism at the university level is handled by harsh punishments, which use fear as a preventative mechanism. Although it is undeniable that academic plagiarism is harmful to students, this approach has multiple shortcomings that may cause more harm than good. A common cause of plagiarism is a student's lack of preparation, indicating that they are not learning at the rate expected by the instructor. Students who are caught plagiarizing often experience a sharp drop in motivation in the class, exacerbating their struggles and fostering a negative attitude towards learning. Due to the ramifications of plagiarism claims, instructors are often hesitant to confront suspected students until they believe there is a strong case to be made. This delay in intervention can be even more harmful as the students continue to deepen their dependence, creating a compounding effect. It is unreasonable to put such burdensome expectations on instructors. Instead of setting up students to fail and instructors to act as police, a system focused on deterrence that works early on, and with as minimal instructor oversight as possible would dramatically improve student outcomes.

The Open Plagiarism Deterrence system (OPD) is presented as a companion to both instructors and students throughout the course of a term. We have designed a 5-stage pipeline to deploy our philosophical approach, as shown in Figure 1. Weekly student code is preprocessed before undergoing pairwise similarity detection and feature generation. A probability measure is then determined from the statistics of the feature distribution, which is presented to the instructional team. Using our custom toolkit, the team may determine weekly warning and action thresholds to be set. Lastly, student independence scores can be distributed to provide feedback on students' measured level of independence. Group and individual statistics can then be monitored in the following week to measure the effectiveness of our deterrence methods.

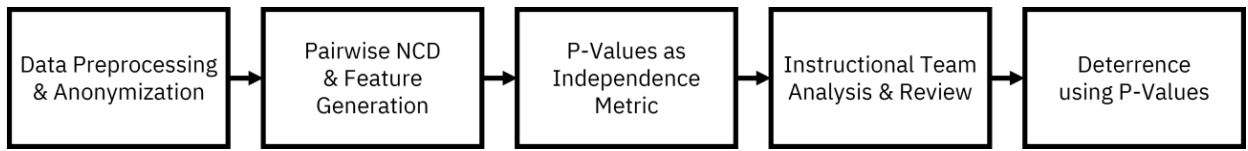

**Figure 1: The Open Plagiarism Deterrence (OPD) system pipeline.** Student submissions are first anonymized. The Normalized Compression Distance (NCD) generates pairwise comparisons. The p-value is formulated from the statistics of the class-wide NCD results. A key characteristic of the p-values is that they sort the submission in order of estimated independence of effort. Guided by that ordering, the instructional team reviews submissions and establishes thresholds for intervention if required.

## Related Literature

The problem of similarity detection is well studied. Most commonly, tools are developed to retroactively detect code similarity between student pairs. The Measure of Software Similarity (MOSS) is well-established and uses a technique called document fingerprinting to compute a precise measure of similarity between two pieces of code [3]. MOSS is the most widely used solution for plagiarism detection. Users execute a Perl script to upload files to the MOSS server, which displays the results on a webpage. Alternatively, research involving compression-based similarity detection has been done to improve plagiarism detection [4, 5]. Both approaches use a lexical analyzer to tokenize student code before constructing pairwise similarity matrices using the Normalized Compression Distance (NCD) [6]. While SID [4] implements a custom compression algorithm, both approaches leverage the existing capabilities of the LZ compressor family to obtain similar results. The system described in [5] opts to use a Shared Nearest Neighbor clustering approach rather than displaying the minimum distanced pairs for instructor review [4]. Both studies ran a range of experiments and report improved performance over tools like MOSS. However, later discussion will demonstrate a drawback of using the produced similarity matrices directly, especially when comparing results across a span of several weeks.

Some effort has been made to implement proactive deterrence mechanisms in programming courses. The Plagiarism Detection Tool (PDT) suggests that real-time deterrence is possible by analyzing individual student progress metrics [7]. By modelling a student's time spent on an assignment, the number of modifications made, and the length of each modification, a classification could in theory be made to directly warn students before they even submit their work. However, this research does not describe an implementation of the proposed system and instead highlights the benefits of prevention compared to retroactive punishment. Similarly, the Temporal Measure of Software Similarity (TMOSS) expands upon the original MOSS implementation by using intermediate snapshots of student code in pairwise comparisons [8]. Although this approach enables instructors to better monitor student collaboration throughout the span of an assignment, there is no mechanism in place to deter students before they submit their work. Yan et al. instead hope that the students' knowledge of this intermediate monitoring would prevent them from engaging in inappropriate collaboration [2, 8].

The problem of plagiarism is complex and requires robust philosophical concepts for an effective solution to be designed. It is now believed that a successful system will integrate powerful detection with appropriate deterrence methods; however, it is often unclear where the line of plagiarism should be drawn [9]. In most cases, plagiarism is non-binary and cannot so easily be identified by an automated system. Furthermore, many non-plagiarism factors can account for some amount of natural similarity in introductory programming courses [10]. While some research has been done to account for this variation and model student similarity, a quantifiable measure of student independence has yet to be established. Important insights can also be gained by understanding why students cheat in the first place. The Fraud Triangle [11] is proposed as an analogy for describing the typical plagiarizer. Given the opportunity and some amount of pressure, a student may find themselves rationalizing an act of plagiarism [2]. This can lead to moral ambiguity and foster negative learning habits within the student body. In order to mitigate confusion regarding what is considered plagiarism, a system should be designed to provide timely and intuitive feedback to both students and instructors.

**Methods**

The first stage of the plagiarism deterrence pipeline involves the extraction, anonymization, and transformation of weekly student code. An anonymizer uses the filenames of the student data to build a secure database of identifying information. Each student is randomly assigned an anonymous ID which is kept track of in the student grading system. At this point, the source data is fully extracted and identifiable only by each student's anonymous ID. The initial anonymization is followed by a more thorough search to ensure no personal information is contained within submitted code. The anonymizer references the student database and searches each file for instances of names, student IDs and email addresses. Additionally, all comments and extraneous newlines are stripped to enhance the compression-based similarity detection discussed later. Anonymization of student submissions enables the subsequent deterrence analysis to be shared transparently with the entire class via a weekly report mechanism.

Following anonymization, each student's submissions are concatenated into a single weekly submission. However, all possible combinations of weekly assignments must be considered to account for students who only submit a portion of them. The power set of assignment submissions is used to identify these different combinations and form corresponding concatenated data. The remaining stages of the system pipeline process each of these submission combinations independently. This preprocessing step is critical for selecting the maximum amount of information to determine a student's perceived likelihood of plagiarism.

The second stage of the system pipeline focuses on extracting similarity features from the weekly concatenated submissions. The normalized compression distance (NCD) [6] is used with the bzip2 compressor to construct pairwise similarity matrices for each submission combination,

$$NCD(x,y) = \frac{C(xy) - \min\{C(x),C(y)\}}{\max\{C(x),C(y)\}}, \qquad (1)$$

where *C(xy)* is the length of the result of compressing the concatenation of submissions *x* and *y*.

Figure 2 shows a subsection of a similarity matrix produced using the NCD. These matrices represent metric distances between students and serve as a legitimate basis for determining code similarity. To better capture similarity relationships in the context of class-wide variance, a new feature, the in-group measure $D_{in}$, was designed based on nearest neighbor relationships between individuals and the group (class),

$$D_{in} = \frac{nearest\ neighbor}{avg.\ distance\ to\ all\ other\ students} \qquad (2)$$

$D_{in}$ identifies pairs of students whose code is especially similar to each other with respect to the average similarity between students and their nearest neighbors. The result is a feature for each student that captures the highest confidence pairwise relationships, normalized to the expected values computed across the entire class. An additional advantage of this feature is that it makes the compression-based similarity measures between weekly analysis more consistent across assignments of increasing complexity.

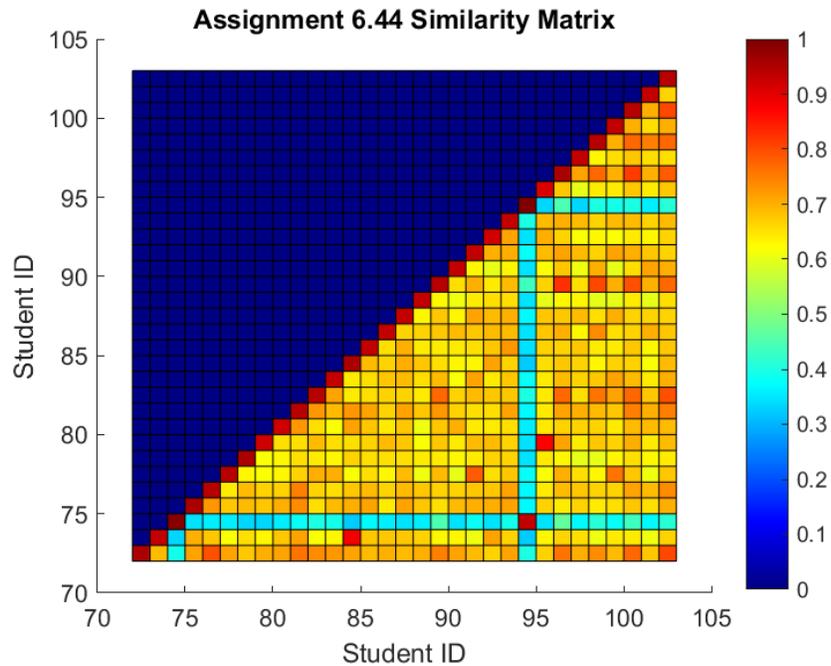

**Figure 1: Visualization of NCD similarity matrix for a previous year's assignment.** The NCD is used to generate pairwise similarities between all students. Note students 74 and 94 at the intersection of the two blue paths. These students are dissimilar from all their peers (as indicated by row and column of light blue), yet highly similar to each other (as indicated by dark red at their intersection). This can often be a sign of using external code from the internet.

The system stores the computed feature distributions in a database for use in subsequent stages. Additionally, spectral plots are generated from the similarity matrices using a t-distributed stochastic neighborhood embedding (TSNE) and stored in a database. Figure 3 shows a comparison of these embeddings against a Laplacian spectral learning approach [12]. Although these plots do not preserve enough meaningful variance for direct use (clustering analysis), they helped form the discussed feature value and serve as at-a-glance validation of group similarity statistics.

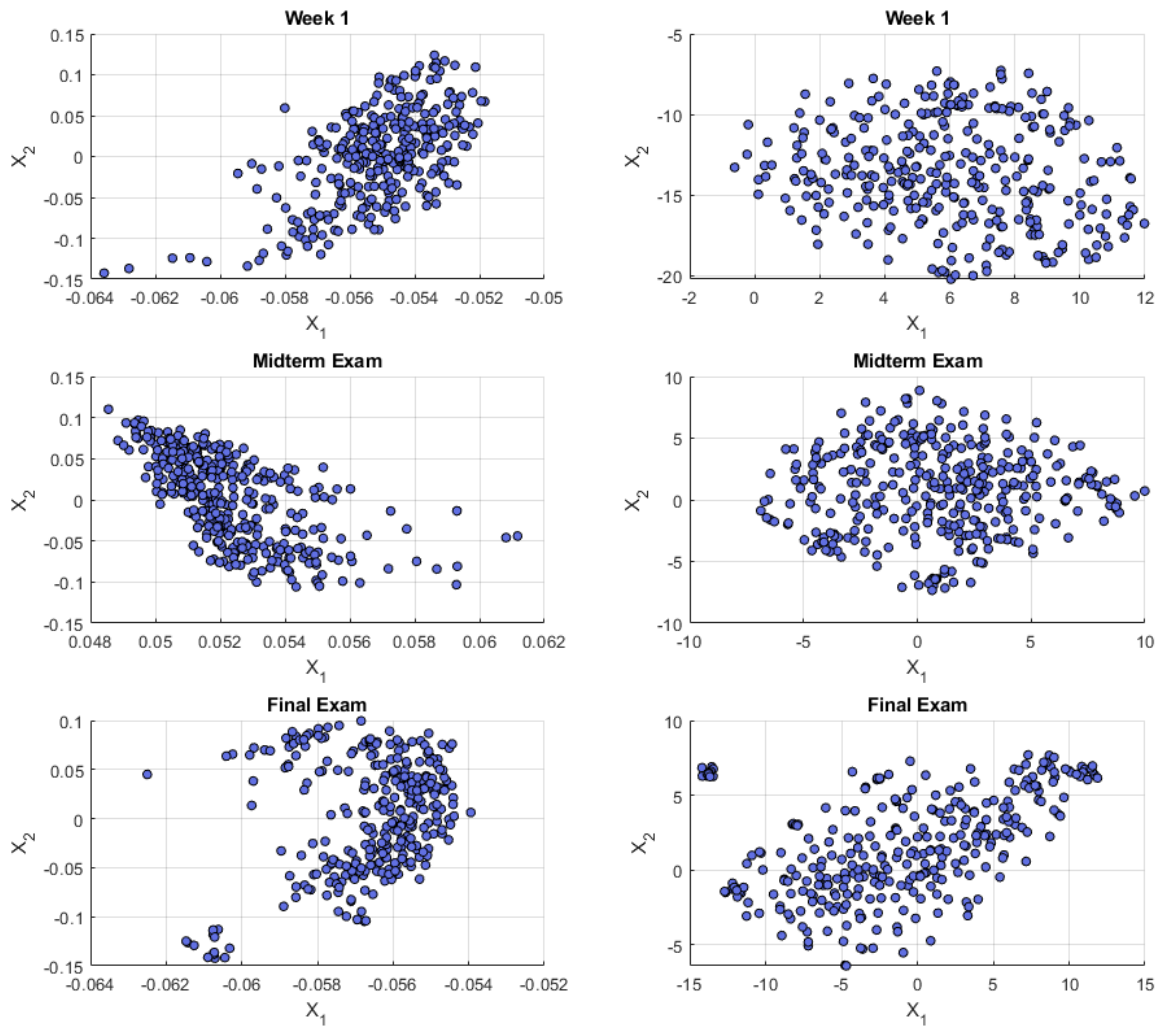

**Figure 2: Weekly Laplacian (left column) and TSNE (right column) embeddings help to visualize student similarity distributions.** Student code is concatenated to form weekly submissions that can be transformed using the NCD and spectral methods [12]. TSNE projections distribute students and isolate clusters better than the Laplacian projections. Although these embeddings are not used to make plagiarism claims, they are key for designing features and validating results.

The empirical feature values are also used to generate an expected distribution. A lognormal distribution is chosen to fit the data based on experimentation. Figure 4 shows how the fitted distribution can be used to isolate high-confidence cases of plagiarism, which allows the instructor to easily select the students with low suspected independence of effort for review.

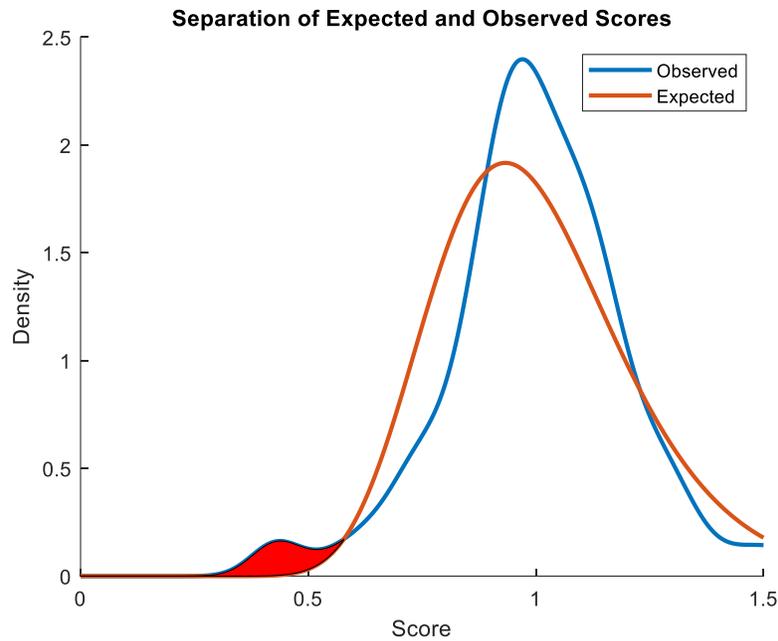

**Figure 3:** The observed curve shows the distribution of nearest neighbor feature scores generated in the previous stage. The expected curve is generated by fitting a lognormal curve to the distribution. **The red region indicates students with exceedingly low independence of effort based on their deviation from the expected distribution.**

In practice, large numbers of students not working independently can bias the distribution. Such unexpected behavior causes these students to appear to have higher independence of effort. Therefore, a filtering stage is used to remove possible outliers from consideration when fitting the expected distribution and preserve the distinction between students working independently versus not. The filtering process simply removes students with scores greater than *k* standard deviations from the mean. The parameter *k* is chosen optimally to minimize an earth-movers distance between the observed and expected distributions.

Once the best fitting expected distribution is computed, the cumulative distribution function (CDF) is used to compute the plagiarism deterrence p-value. The p-values can be interpreted as the probability that the student worked independently from their nearest neighbor. This independence of effort p-value is the essential output of the OPD system. First, it allows the submissions to be sorted in order of increasing independence, since the p-values correspond monotonically with the feature scores. This behavior enables efficient instructor review of results in the final stages of the pipeline. Second, the simple translation of feature scores to p-values using the CDF makes the system transparent. Many existing plagiarism systems, such as MOSS, use complicated black-box methods that obscure the internal processing. The design of this system allows instructors to view and understand how a result is computed. Throughout the experiment, the instructional team reviewed figures such as Figure 5 to understand the distribution of student feature scores. Finally, the deterrence p-values can be intuitively understood by students. The p-values can be directly distributed to students along with the assignment grade, communicating a suspected level of independence of effort. Each component of the OPD system works to maintain consistency among p-values across all assignments in a term. As a result, p-values very closely approximate the true instructor confidence of the independence of student effort for every assignment. Additionally, assigning weekly independence scores to students reduces ambiguity regarding what is considered plagiarism. Clear and timely feedback enables instructors to act earlier than previously possible so that positive student habits can be formed.

Once p-values have been computed for each student, the instructor can begin to review the results. Weekly results are made available through a comprehensive web application and autogenerated reports. The reports list submissions in order of increasing independence. For each submission, the report includes a code diff of the submissions, the nearest neighbor pair's p-value, and a chart showing the expected and observed distributions with the pair's score marked on the chart (Figure 5). TSNE spectral plots are also included for reference (Figure 3).

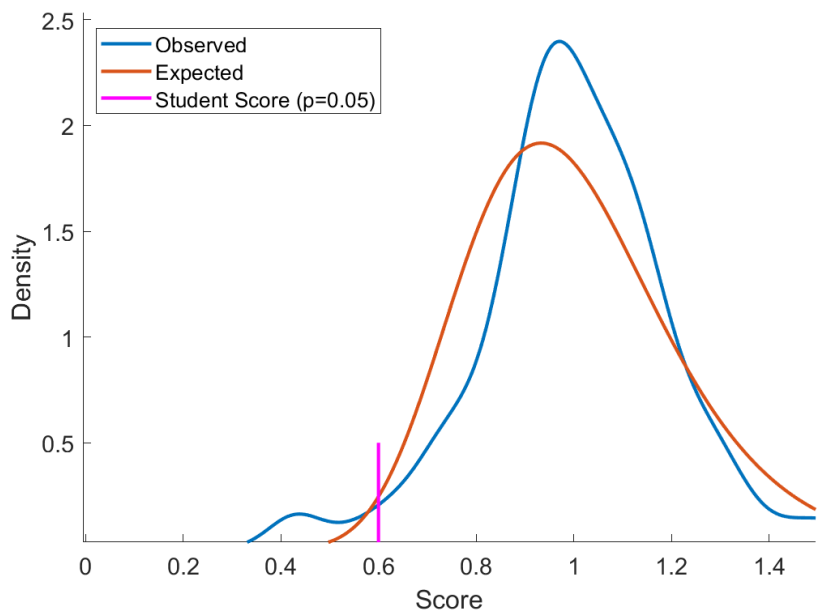

**Figure 4: Sample report figure showing a particular student's score in context with the rest of the class. The distributions can help inform an instructor's decision, by showing where separation occurs.**

The instructional team then searches through the results until they reach the first submission where they are no longer highly confident of plagiarism. At that point, a threshold can be chosen. For the weekly assignments included as part of our open data, this threshold falls around $p \leq 0.01$. **Without checking all assignments, the instructor can feel confident that there are no high-confidence cases with p-values higher and no low-confidence cases with p-values lower than the chosen threshold.** This point is referred to as the action threshold. Below the action threshold, the instructional team feels confident taking disciplinary action, such as giving a failing grade for the assignment, or reporting the offending students for academic integrity violations. The second threshold, called the warning threshold, is used for deterrence purposes. This threshold indicates to students who fall between it and the action threshold that they are dangerously close to being in clear violation. These are lower-confidence cases, where gentle deterrence is a much more effective solution to improve student outcomes than direct action.

While instructors may set the warning threshold on a per-assignment basis, they may consider instead using a static threshold to avoid confusing students. Since p-values measure independence of effort, they can be used across assignments of varying complexity and length. Additionally, this method of instructor review and thresholding greatly reduces the chance of a false positive occurring. The cost of a false plagiarism claim is very high, as it can disrupt trust between students and instructors and even have greater implications with respect to the institution. As such, the OPD system provides the instructional team with the proper information to make informed plagiarism claims.

The instructional team can also view historical patterns between a given pair of students using the web application provided in the open-source toolkit. This feature shows p-values for all previous assignments between a pair of students. If the pair had a history of low p-values, the instructor would be more inclined to treat them as a high-confidence case, since students that cheat together have been shown to do so on several occasions [4].

To facilitate deterrence, p-values are directly distributed to students. For this research, the instructional team opted to upload them as a 'grade' into the institution's e-learning platform. Doing so allows students to directly access results of the deterrence system, while maintaining confidentiality. Students falling below the action threshold were contacted directly by the instructional team for intervention and disciplinary action. Within the warning threshold, the responsibility was on the student to check their p-value score themselves. Conscientious students would notice if they scored within or close to the warning threshold. This form of deterrence is entirely self-directed, which avoids a confrontation that can discourage and dishearten students.

The OPD system is designed for consistent, weekly use by the instructor. Unlike existing tools, our pipeline serves as a course companion that enables validation of the described deterrence mechanism. By analyzing results across many weeks, the instructor can better gauge both individual and group independence performance.

**Results and Discussion**

The combination of stages described above form the core of an effective plagiarism detection and deterrence system. During the 2022 Winter term, the system was deployed live in an introductory programming course with the intent to validate its detection capabilities and test the efficacy of the chosen deterrence method. Students were informed from the beginning that their work would be subject to an experimental plagiarism detection software, and that its results would be closely reviewed by the instructional team. The course had 425 students. The Python language was used, but the approach will work with any language with some minor modifications to the system. Student submissions consisted of several weekly independent assignments. In addition, students were assigned a two-part midterm and two-part final exam in a similar format to their weekly homework. Each week, student submissions were processed by the OPD system. Weekly results were then reviewed by the instructional team before deterrence was deployed via p-values. Student p-values were closely monitored each following week to gauge the effectiveness of the deterrence mechanism. Additionally, results of the ODP system were often compared to the built-in zyBooks similarity checker [13].

Due to the trial-like nature of the experiment, there were inconsistencies with how deterrence and instructor intervention were applied. The instructional team was hesitant, especially in earlier weeks, to rely on the results, even with manual review. Often, the team would review a pair of submissions that were near-identical and struggle to say concretely whether such identicality was coincidental, even with p-values below 1e-5 (99.999% confidence). As the experiment progressed and student submissions became more sophisticated, the team began to learn how to interpret the results. By the midterm exam, the team began distributing p-values and high-confidence interventions. Out of 369 submissions, the team identified seven students below the determined action threshold. These students received zero credit for the exam and were offered the opportunity to contest. None of the students disputed the results, and two confessed to cheating. In addition to direct interventions, the team posted p-values on the university's e-learning platform to begin monitoring the impact on future student levels of independence.

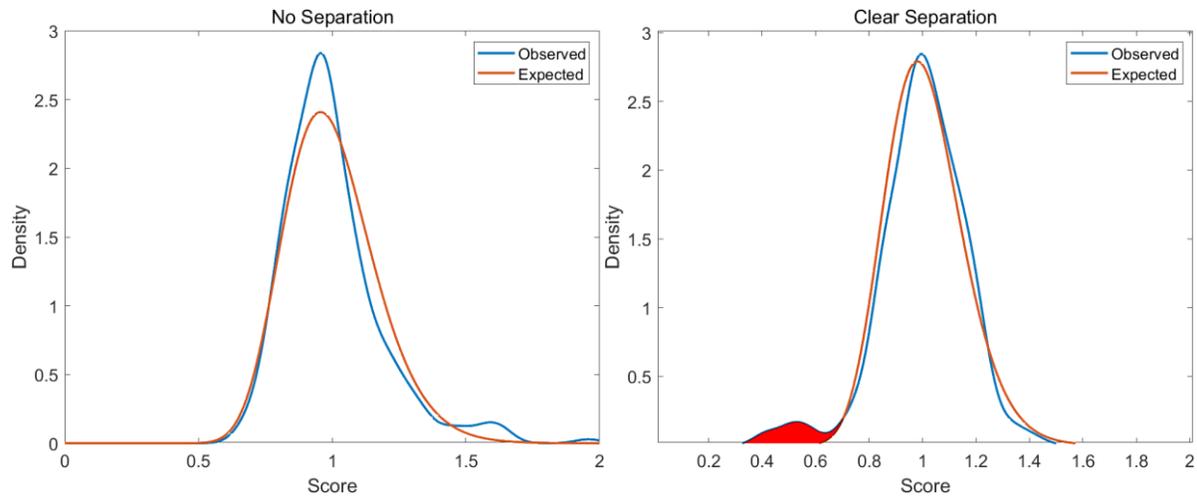

**Figure 5:** (Left) Weeks with no cheating produce curves with no separation. No students had outstandingly low scores. (Right) Red region highlights region of high-confidence. **Separation demonstrates that students falling into the red region have distinctly low distances outside of the expected range.** Right-edge of red region corresponds to approximately $p = 0.05$.

Although students were aware that their submissions were being monitored by a plagiarism detection system, the midterm was the first time that they were notified of the results. To address the large number of students concerned by their low p-value scores, the team devised some guidance on how students can demonstrate their independence of effort in a way that would minimize the chances of them being detected by the system. It was suggested to maximize the independence of work by implementing the same code path using different mechanisms, e. g. looping by iterating over structures versus iterating via range objects. The additional variability introduced in this way raises the expected nearest neighbor feature score for the class and creates greater variance, allowing easier separation of cheating and non-cheating students. In a distribution of scores with no cheating, the expected value curve will nearly perfectly align to the observed dataset, with no students having p-values below 0.05 (Figure 6).

When inspecting the results from the final exam, it became exceedingly clear that the presence of many highly similar submissions could heavily bias the model (Figure 7). To address this, the filtering method described earlier was applied to normalize p-values in these cases. Applying this method retroactively to the previous weeks considerably improved separation in all cases. Additionally, it normalized student p-values across all weeks.

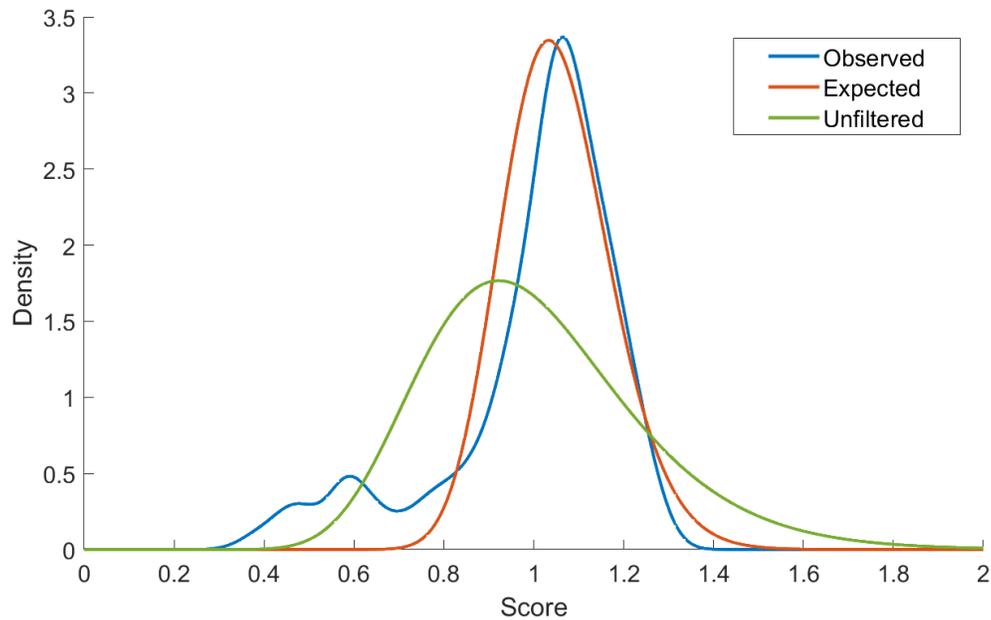

**Figure 7: Large amounts of cheating on the final exam caused the expected curve (green) to be flattened and shifted left. This results in high p-values in cases of high-confidence plagiarism. To account for this type of result, filtering is applied to push the expected curve (orange) as close to the observed curve (blue) as possible.** Since the expected curve should model the class with no cheating, the focus is to identify where scores start to substantially deviate from the mode of the class. Students are removed in increasing order of score to achieve the match.

Alternative, more aggressive deterrence mechanisms were originally considered. An automated email routine was developed in conjunction with a sophisticated escalation system (both available in the open-source toolkit as optional modules) to warn students of repeated low-confidence detections and high-confidence instances of plagiarism. However, the instructional team found that this approach would appear too intimidating and impersonal to the students. Such a direct approach was considered antithetical to the core philosophy of early deterrence for encouraging independence. Instead, the team opted to take advantage of the intuitiveness of p-values by sharing them directly with students. Sharing the results with students improves transparency and trust, focusing more on the deterrence, rather than the detection aspects of the system. Students may view their p-values for any given week and respond accordingly. For example, a student working with a peer might see that they received a low p-value. This pair would now know that they were working too closely and should focus on working more independently. Avoiding confrontation also reduces the harm associated with a false positive, as students feel less threatened, and instructors are free to use the tool with less rigorous supervision.

```
29  def countBases(seq):                                          29  def countBases(seq):
30      A=seq.count('A')                                          30      A=seq.count('A')
31      C=seq.count('C')                                          31      C=seq.count('C')
32      G=seq.count('G')                                          32      G=seq.count('G')
33      T=seq.count('T')                                          33      T=seq.count('T')
34      seq_1= C+G                                                34      yuh=C + G
35      seq_2= C+G+T+A                                            35      hi=C+G+T+A
36      cg_percentage = (seq_1/seq_2)*100                         36      cg_percentage = (yuh/hi)*100
37      return '{:d} {:d} {:d} {:d}\n{:2.1f}'.format(A, C, G, T, cg_percentage)   37      return('{:d} {:d} {:d} {:d}\n{:2.1f}'.format(A, C, G, T, cg_percentage))
38  def hammingdistance(seq_a, seq_b):                            38  def hammingdistance(seq_a, seq_b):
39      if (len(seq_a) !=len(seq_b)):                             39      if len(seq_a)!=len(seq_b):
40          return ('Error: sequences length mismatch')           40          return('Error: Sequences Length Mismatch')
41      else:                                                     41      else:
42          seq_a=set(seq_a)                                      42          seq_a=set(seq_a)
43          seq_b=set(seq_b)                                      43          seq_b=set(seq_b)
44          seq=seq_a.symmetric_difference(seq_b)                 44          ekk=seq_a.symmetric_difference(seq_b)
45          seq=set(seq)                                          45          ekk=set(ekk)
46          seq=len(seq)                                          46          ekk=len(ekk)
47          return 7                                              47          return 7
```

**Figure 8:** Sample diff showing partial submission for midterm exam. Note the variable name changes in lines 34 and 35. **These were clearly made by a student attempting to thwart detection. Such changes had no impact, and the OPD rated the pair with a p-value of 2.01e-3, well within the action threshold.**

Figure 8 shows a code-diff between partial submissions from two students' midterm exams. These two submissions were almost identical except for this section. Clearly, the student on the right copied from the student on the left and lazily changed some variable names to avoid detection. However, the NCD is robust to code rearrangement and variable renaming [6]. The OPD assigned this pair $p = 2.01\text{e-}3$, well within the high-confidence range. For this submission, the inbuilt zyBooks plagiarism detection buried these students under hundreds of other submissions. They received a score of only 7.5 out of 10 and would likely go undetected unless the instructor chose to comb through the hundreds of other matches above them.

One major advantage the OPD has over other systems comes from the feature design limiting the number of comparisons to be considered. Our approach constructs a single similarity score per student ($O(n)$), where existing approaches such as MOSS and zyBooks consider all pairwise distances ($O(n^2)$). In addition to searching through hundreds of comparisons, another frustration the team had with zyBooks was that for most assignments, it would rate too many comparisons at the maximum similarity score. This is likely due to a problem seen in most similarity checkers, where they use a preprocessing stage to remove certain information to optimize detection. For example, zyBooks' system removes all variable names so that they are not considered [13]. These machines are better suited for more complicated code written by more sophisticated programmers. In an introductory class, relatively simple changes in flow and ordering can be enough to show that two programmers are working independently. However, the experimental data suggests that most students who are cheating are not experienced enough to make these changes. The NCD detects shared variable names and similar changes with great precision, and scores submissions accordingly. The reports generated by the OPD system arrange these scores monotonically, making the process of thresholding easy. This reduces the number of comparisons instructors must review.

Figure 9 shows the weekly distributions of p-values. The p-values were first shared during week six, following the midterm exam in week five. This allowed us to validate the performance of the p-value in ranking and identifying independence of programming effort. There is an increase in independence of effort during week six, following the first release of p-values. Week seven contained one of the more difficult assignments of the quarter (matplotlib and NumPy), with few degrees of freedom, which lead to a decrease in p-values. In future experiments, p-values will be shared starting in week one.

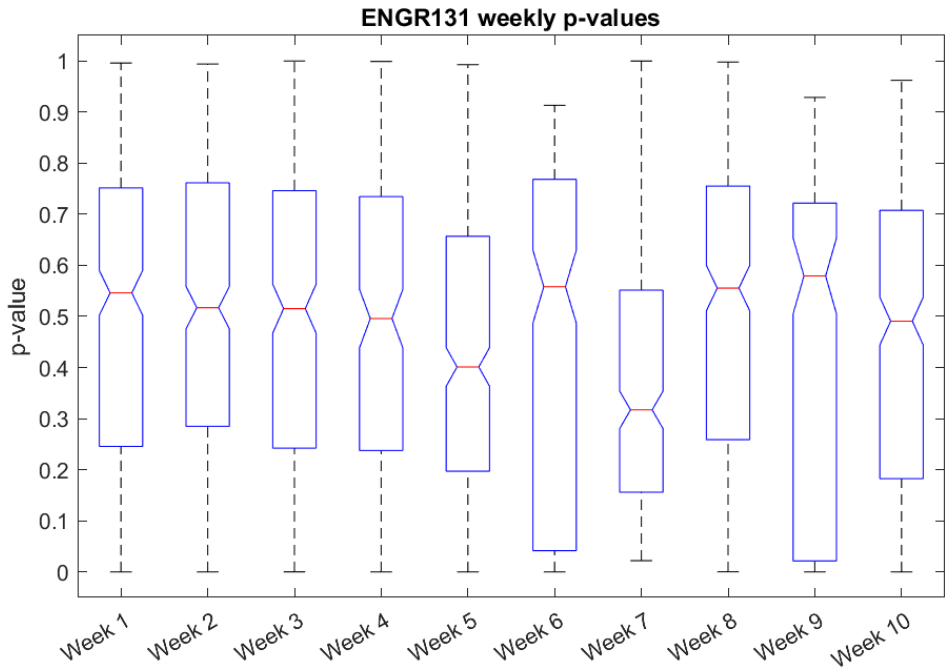

**Figure 9:** Distributions of p-values weekly through a ten-week term. The blue box represents 25% and 75% of the values, the red line is the median. The whiskers show outliers. The p-values were first made available to students during week six with the midterm exam results from week five. Note the strong decrease in p-values during week seven where a particularly challenging set of problems were assigned.

## Source Code and Experimental Data

All source code from this project, along with the anonymized code submissions and weekly summary reports, are available at https://git-bioimage.coe.drexel.edu/opensource/plagiarism_deterrence/-/tree/main. The source code is released under the MIT license.

## Conclusion

One concern with directly exposing the results of the system without any post-processing is that it allows students to attempt to outsmart OPD. Fortunately, at the introductory level, the types of changes a student might make to avoid detection require a fundamental understanding of coding concepts. Spacing, comments, variable name and style changes are not enough to thwart detection. Instead, the code must be restructured to implement different logic.

The Open Plagiarism Deterrence system is a set of mathematical approaches, software tools, and labeled sampled data (code submissions). The goal of the system is to enable the measurement of independence of effort in an ensemble of student submissions for beginning programming courses. In doing so, the system could provide early encouragement for students to learn to code independently. The system was effective even from week one, with very simple programming assignments. Using the statistical characteristics of the whole class submissions enables a comprehensive characterization of individual efforts.